\documentclass[a4paper,11pt]{article}
\usepackage{pos}
\usepackage{hyperref}

\title{\mimes, the Misalignment Mechanism Solver}

\author{Dimitrios Karamitros}

\affiliation{School of Physics and Astronomy, The University of Manchester,\\ Manchester M13 9PL,
	United Kingdom}

\emailAdd{dimitrios.karamitros@manchester.ac.uk}

\abstract{We briefly discuss \mimes, a \CPP library that can solve the axion and axion-like equation of motion in various axion models and cosmological histories.}

\FullConference{%
  Computational Tools for High Energy Physics and Cosmology (CompTools2021)\\
  22-26 November 2021\\
  Institut de Physique des 2 Infinis (IP2I), Lyon, France
}

\usepackage{ifthen}
\usepackage{tikz}
\usepackage{xspace}
\usepackage{listings}
\usepackage[T1]{fontenc}

\definecolor{mediumjunglegreen}{rgb}{0.11, 0.21, 0.18}
\definecolor{bg}{HTML}{282828}

\newcommand*\cppin{\lstinline[language=c++]}

\newcommand*\pyin{\lstinline[language=c++]}

\lstnewenvironment{cpp}
{\lstset{language=c++}}
{}

\lstnewenvironment{py}
{\lstset{language=Python}}
{}

\lstnewenvironment{bash}
{\lstset{language=Python}}
{}

\lstnewenvironment{pseudo}
{\lstset{language=Python}}
{}

\newcommand{\ie}{{\em i.e.}\xspace}
\newcommand{\eg}{{\em e.g.}\xspace}
\newcommand{\GeV}{{\rm GeV}\xspace}

\def\mimes{{\tt MiMeS}\xspace}

\newcommand{\CPP}{{\tt C++}\xspace}

\newcommand{\PY}{{\tt python}\xspace}

\newcommand{\rhs}{RHS\xspace}

\newcommand{\thetai}{ \theta_{\rm ini}{}\xspace}

\newcommand{\fa}{ f_{a}{}\xspace}

\newcommand{\ma}{ m_{a}{}\xspace}
\newcommand{\maT}{ \tilde m_{a}{}\xspace}

\newcommand{\lrb}[1]{\left( #1 \right)}

\newcounter{NumArgs}

\newcommand{\eqs}[1]{\setcounter{NumArgs}{0}\foreach\i in{#1}{\stepcounter{NumArgs}}%
	\ifthenelse{\equal{\theNumArgs}{1}}{eq.~(\ref{#1})}%
	{\ifthenelse{\equal{\theNumArgs}{2}}%
		{eqs.~\foreach\i[count=\q]in{#1}{\ifthenelse{\equal{\q}{\theNumArgs}}{and (\ref{\i})}{(\ref{\i})~}}}%
		{eqs.~\foreach\i[count=\q]in{#1}{\ifthenelse{\equal{\q}{\theNumArgs}}{and (\ref{\i})}{(\ref{\i}),~}}}}}

\newcommand{\Eqs}[1]{\setcounter{NumArgs}{0}\foreach\i in{#1}{\stepcounter{NumArgs}}%
	\ifthenelse{\equal{\theNumArgs}{1}}{Eq.~(\ref{#1})}%
	{\ifthenelse{\equal{\theNumArgs}{2}}%
		{Eqs.~\foreach\i[count=\q]in{#1}{\ifthenelse{\equal{\q}{\theNumArgs}}{and (\ref{\i})}{(\ref{\i})~}}}%
		{Eqs.~\foreach\i[count=\q]in{#1}{\ifthenelse{\equal{\q}{\theNumArgs}}{and (\ref{\i})}{(\ref{\i}),~}}}}}

\newcommand{\refs}[1]{\setcounter{NumArgs}{0}\foreach\i in{#1}{\stepcounter{NumArgs}}%
	\ifthenelse{\equal{\theNumArgs}{1}}{(\ref{#1})}%
	{\ifthenelse{\equal{\theNumArgs}{2}}%
		{\foreach\i[count=\q]in{#1}{\ifthenelse{\equal{\q}{\theNumArgs}}{and (\ref{\i})}{(\ref{\i})~}}}%
		{\foreach\i[count=\q]in{#1}{\ifthenelse{\equal{\q}{\theNumArgs}}{and (\ref{\i})}{(\ref{\i}),~}}}}}

\newcommand{\Figs}[1]{\setcounter{NumArgs}{0}\foreach\i in{#1}{\stepcounter{NumArgs}}%
	\ifthenelse{\equal{\theNumArgs}{1}}{Fig.~(\ref{#1})}%
	{\ifthenelse{\equal{\theNumArgs}{2}}%
		{Figs.~\foreach\i[count=\q]in{#1}{\ifthenelse{\equal{\q}{\theNumArgs}}{and (\ref{\i})}{(\ref{\i})~}}}%
		{Figs.~\foreach\i[count=\q]in{#1}{\ifthenelse{\equal{\q}{\theNumArgs}}{and (\ref{\i})}{(\ref{\i}),~}}}}}

\newcommand{\Gen}[2]{\setcounter{NumArgs}{0}\foreach\i in{#2}{\stepcounter{NumArgs}}%
	\ifthenelse{\equal{\theNumArgs}{1}}{#1.~(\ref{#2})}%
	{\ifthenelse{\equal{\theNumArgs}{2}}%
		{#1.~\foreach\i[count=\q]in{#2}{\ifthenelse{\equal{\q}{\theNumArgs}}{and (\ref{\i})}{(\ref{\i})~}}}%
		{#1.~\foreach\i[count=\q]in{#2}{\ifthenelse{\equal{\q}{\theNumArgs}}{and (\ref{\i})}{(\ref{\i}),~}}}}}

\begin{document}
\maketitle

\section{Motivation}
	One class of promising dark matter candidates are the Axions~\cite{Peccei:1977hh,Weinberg:1977ma,Wilczek:1977pj,Preskill:1982cy,Dine:1982ah,Abbott:1982af,Berezhiani:1989fp,Berezhiani:1992rk,Sakharov:1994id,Sakharov:1996xg,Khlopov:1999tm} as well as axion-like paricles (ALPs)~\cite{Chikashige:1980ui,Georgi:1981pu,Ringwald:2014vqa}, which follow an equation of motion (EOM)
	\begin{equation}
		\lrb{\dfrac{d^2}{d t^2} + 3 H(t) \ \dfrac{d}{d t} } \theta(t) + \maT^2(t) \ \sin \theta(t) = 0 \; ,
		\label{eq:eom}
	\end{equation}	
	where $\theta= A \, \fa$, with $A$ the axion field, and $\fa$ some energy scale that characterises the potential (Peccei-Quinn breaking scale). 
	The classical analogue is the dumped pendulum with both frequency (length) and friction being time-dependent. That is, there is no closed form solution, 
	no constants of motion until the Hamiltonian starts to vary slowly.

	In this proceedings, we present the basic aspects of \mimes~\cite{Karamitros:2021nxi}. \mimes is designed to solve the EOM~(\ref{eq:eom}) numerically; \ie~\mimes simulates the evolution of the axion/ALP, for many cosmological scenarios and axion/ALP (thermal) masses. In short, \mimes is a \CPP header-only library that contains various templated classes; there is no ``installation" and no special procedures, just include the header files. It comes with a \PY interface making it accessible to everyone. \mimes also is easy to use; anyone can run it and see if their model can work or check against the literature. Moreover, the user is 
	free to decide when to start, stop, and when adiabaticity is reached. So, almost every aspect of the numerical solution can be tuned by the user.

\section{About \mimes}
	\subsection{Under the hood}
	\mimes relies on \cppin{NaBBODES}~\cite{NaBBODES} for the numerical integration, and \cppin{SimpleSplines}~\cite{SimpleSplines} for the various interpolations. 
	These are header-only libraries developed and maintained by the author. They are included in the source code of \mimes, which means that \mimes only needs the standard \CPP library.
	Moreover, their integration with \mimes is guaranteed, since there will always be suitable versions of these libraries that work with the current version of   \mimes.
	
	\subsection{How to get \mimes}
	There are several ways one can get a stable version of \mimes. First, via github, by running \cppin{git clone -b stable https://github.com/dkaramit/MiMeS.git}.
	This is the preferred way, as it is guaranteed to be the latest stable version. Moreover, a copy can be downloaded from 
	\href{https://mimes.hepforge.org/downloads}{mimes.hepforge.org/downloads} or \href{https://github.com/dkaramit/MiMeS/releases}{github.com/dkaramit/MiMeS/releases}.
	In order to get the most up-to-date code -- not always the most stable one -- including the latest version of {\tt NaBBODES} and {\tt SimpleSplines}, one needs to run
	the following commands
	\begin{lstlisting}
		git clone https://github.com/dkaramit/MiMeS.git
		cd MiMeS
		git submodule init
		git submodule update --remote
	\end{lstlisting}

	\subsection{Configure (and make)}
	If the user intents to use \mimes directly from \CPP they do not need to install anything in particular. The only requirement is to run 
	\begin{bash}
		bash configure.sh
	\end{bash}
	in the root directory of \mimes. After that, the classes and functions of \mimes can be used by including the header file \cppin{MiMeS/MiMeS.hpp}.
	However, there are several examples in \pyin{MiMeS/UserSpace/Cpp}, which can be modified in order to meet the needs of the user. These examples can be 
	compiled by running  \cppin{make examples} in the root directory of \mimes, or simply \cppin{make} in the sub directories inside \pyin{MiMeS/UserSpace/Cpp}.
	
	In order to call \mimes  from \PY scripts, one needs to compile the shared libraries that translate the \CPP classes to {\tt C}-style functions that can be loaded to 
	\PY. This can be done simply by running \cppin{make lib} in the root directory of \mimes.

	\subsection{Classes}
	There are three classes useful to the user. The first one that is responsible for the interpolation of relativistic degrees of freedom of the plasma, \cppin{mimes::Cosmo<LD>}.
	Although this is called internally  by others, the user should be aware of its existence, since the file that contains the relevant data (by default, these are taken from~~\cite{Saikawa:2020swg})  can be changed freely. The class used to define the axion mass as a function of $\fa$ and the temperature is \cppin{mimes::AxionMass<LD>}.  
	\mimes comes with data from Lattice calculation~\cite{Borsanyi:2016ksw} of the QCD axion mass, but the user can use other data or a simple function.
	The class that is responsible for actually solving the EOM is \cppin{mimes::Axion<LD,Solver,Method>}. Here, the user is free to use a number of different Runge-Kutta methods,
	as well control various aspects of how integration happens. The details on how exactly these classes work can be found in ref.~\cite{Karamitros:2021nxi}.
	
	All classes take a template argument labelled \cppin{LD}, which should be \cppin{double} (fast) or \cppin{long double} (accurate). For most cases, the latter is
	recommended, since $\theta$ can be very small especially in cosmologies with entropy injection. Moreover, the other template arguments are
	\begin{itemize}
		\item \cppin{Solver} can be set to \cppin{1} for Rosenbrock (semi-implicit Runge-Kutta). 
		The \cppin{Method} argument in this case can be:~\footnote{We should note that the user is free to build their own RK method, as illustrated in~\cite{Karamitros:2021nxi}.}
		\begin{itemize}
			\item \cppin{RODASPR2<LD>} (4th order). This is the most accurate one, and the one that is used in all the examples.
			\item \cppin{ROS34PW2<LD>} (3rd order). This is generally less accurate, but it can be used for quick estimates.
			\item \cppin{ROS3W<LD>} (2rd order). This, generally should be avoided since it is inaccurate.
		\end{itemize}
		\item \cppin{Solver} can be set to \cppin{2} for explicit RK.
		The \cppin{Method} argument can be:
		\begin{itemize}
			\item \cppin{DormandPrince<LD>} (7th order). This is an explicit RK, and not very good for this EOM. However, it can still be used.
			\item \cppin{CashKarpRK45<LD>} (5th order). This method should be avoided as it fails to converge in most cases.
			\item \cppin{RK45<LD>} (5th order). Similarly to the previous one, this also seems to have difficulty converging.
		\end{itemize}	
	\end{itemize}

	\subsection{Calling \mimes from \PY}
	In order to call the \PY interface of \mimes, we need to first call \cppin{make lib} in the root directory of \mimes. This compiles the shared libraries 
	that \PY needs. Before that, the user can change the template arguments and various compilation options. This can be done by changing the various variables in the 
	file \cppin{MiMeS/Definitions.mk} 
	\begin{itemize}
		\item \cppin{LONGpy=long} will compile the library with \cppin{long double} numeric types. \cppin{LONGpy=} will compile the library with \cppin{double} numeric types.
		\item \cppin{SOLVER} and \cppin{METHOD}, are in the template arguments for the \CPP case.
		\item Compiler:
		\begin{itemize}
			\item \cppin{CC=g++} in order to use the \cppin{GNU} \CPP compiler. This is the compiler used to test \mimes; \ie should be the preferred choice.
			\item \cppin{CC=clang -lstdc++} in order to use the \cppin{clang} \CPP compiler. Sometimes this is faster than {\tt g++}, but should be used carefully.
		\end{itemize}
		\item Optimization level:
		\begin{itemize}
			\item \cppin{OPT=O0}: No optimization. The executables compiled using this option are generally slow compared to the following choices.
			\item \cppin{O=O1}, \cppin{O2}, or \cppin{O3}: all these perform mostly the same (read the compiler documentation for more information on the optimization). Generally, the latter
			is preferred.
			\item \cppin{OPT=Ofast}: full optimization. This choice produces the fastest executables. Although it is generally considered dangerous, we have not observed any
			negative side-effects.  
		\end{itemize}
	\end{itemize}

	\subsection{Assumptions}
	\mimes is designed to make as few assumptions as possible. The basic assumption it makes are:
	\begin{enumerate}
		\item $\dot \theta(0)=0$. This means that at high temperatures the kinetic energy dominates, $\dot \theta$ is a conserved current; \ie it falls as $a^{-3}$ ($a$ is the scale factor). Therefore, at low temperatures, $\dot \theta(0)$ can be assumed to be negligible.   
		\item $H/\maT$ increases monotonically with the temperature. This is related to the previous assumption. Practically, it is also used to find an appropriate starting integration point. 
		\item The energy density of the axion/ALP is always subdominant, since the axion/ALP is assumed to be at least a DM component. This ensures that the axion/ALP evolves in some cosmological background. Otherwise, one would need to solve a system of equations that take into account the effect of the axion to the evolution of the rest of Universe (\eg contribution to plasma energy density).
		\item Only \eqs{eq:eom} determines the axion/ALP energy density. That is, there are other sources of axion production.
	\end{enumerate}

\section{Examples}
	There are several examples in both \CPP and \PY that can be found in \cppin{MiMeS/UserSpace}. However, it would be useful to append two simple ones here.
	
	\subsection{\CPP example}
	In this example, we show how to solve the QCD axion EOM for a matter dominated universe, using $\thetai=0.1$ and $\fa=10^{16}~\GeV$. The data for the mass used is given in~\cite{Borsanyi:2016ksw}. The path of this file is stored by in the global variable \cppin{chi_PATH} that is created when we run the \cppin{configure.sh} script.	The evolution of the Hubble parameter is solved separately (see \eg ref.~\cite{Arias:2020qty}), and the corresponding file is \cppin{MiMeS/UserSpace/InputExamples/MatterInput.dat}.   
	
	\lstset{language = c++}
\begin{lstlisting}[basicstyle=\footnotesize]
#include<iomanip>
#include"MiMeS.hpp"

using numeric = long double;//make life easier if you want to change to double

int main(){
    mimes::util::Timer _timer_;//use this to time it!

    // use chi_PATH to interpolate the axion mass.
    mimes::AxionMass<numeric> axionMass(chi_PATH,0,mimes::Cosmo<numeric>::mP);

    /*set ?$\maT^2$? for ?$T\geq T_{\rm max}$?*/
    numeric TMax=axionMass.getTMax(), chiMax=axionMass.getChiMax();    

    axionMass.set_ma2_MAX(
        [&chiMax,&TMax](numeric T, numeric fa){ return chiMax/fa/fa*std::pow(T/TMax,-8.16);}
    );  
    /*set ?$\maT^2$? for ?$T\leq T_{\rm min}$?*/
    numeric TMin=axionMass.getTMin(), chiMin=axionMass.getChiMin(); 

    axionMass.set_ma2_MIN( 
        [&chiMin,&TMin](numeric T, numeric fa){ return chiMin/fa/fa;}
    );		

    /*this path contains the cosmology*/
    std::string inputFile = std::string(rootDir)+
        std::string("/UserSpace/InputExamples/MatterInput.dat");

    /*declare an instance of Axion*/
    mimes::Axion<numeric, 1, RODASPR2<numeric> > ax(0.1, 1e16, 500, 1e-4, 1e3, 10, 1e-2, 
                    inputFile, &axionMass, 1e-2, 1e-8, 1e-2, 1e-10, 1e-10, 0.85, 1.5, 0.85,
                    int(1e7) );
    /*solve the EOM!*/
    ax.solveAxion();

    std::cout<<std::setprecision(5)
    <<"theta_i="<<ax.theta_i<<std::setw(25)<<"f_a="<<ax.fa<<" GeV\n"
    <<"theta_osc~="<<ax.theta_osc <<std::setw(20)
    <<"T_osc~="<<ax.T_osc<<"GeV \n"<<"Omega h^2="<<ax.relic<<"\n";
    
    return 0;
    }
\end{lstlisting}

	We note here that the axion mass is interpolated between the temperatures that exist in the corresponding file. Beyond these points, we set the mass to follow the functions
	that are passed as arguments in \cppin{set_ma2_MAX} and \cppin{set_ma2_MIN}.  Here, we choose
	\begin{equation}
		\maT = \Bigg\{
		\begin{matrix}
			\dfrac{\chi(T_{\rm min})}{\fa^2} & \text{for } T<T_{\rm min} 
			\\ \\
			\dfrac{\chi(T_{\rm max})}{\fa^2}   \lrb{\dfrac{T}{T_{\rm max}}}^{-8.16} & \text{for } T>T_{\rm max} 
		\end{matrix} \;.
		\label{eq:axM-limits}
	\end{equation}
	Once the mass is defined according to our model, we move to setting-up and solving the EOM. This is done in lines $30$-$34$. Observe that the constructor of the \cppin{mimes::Axion} class takes a lot of arguments. The role of all these arguments is explained in detail in the documentation of \mimes~\cite{Karamitros:2021nxi}. However, it is important to note that we choose as a starting point the temperature where $H/\ma = 1000$. One can detemine if this is a valid value by taking a larger one and confirming that the end result does not change.
	 Also, we consider the axion to start evolving adiabatically at the point where the relative difference of its adiabatic invariant is less that $10^{-2}$ between peaks of the oscillation for $10$ consecutive times.
	 
	 Assuming that the code is saved in a subdirectory of the root of \mimes, we can compile this code as 
	 \begin{bash}
	 	g++ -O3 -std=c++17 -lm -I../ -o axion example.cpp
	 \end{bash}
	Then, the executable (\cppin{axion}) can be called as \cppin{./axion}.

	\subsection{\PY example}
	\mimes can also be called from \PY using a script with similar structure. The \PY interface is made as close to the \CPP code as possible. This can be seen by the following example, where 
	we show a \PY script that performs the same calculation as the \CPP example, in a very similar way.
	
	\lstset{language = python}
\begin{lstlisting}[basicstyle=\footnotesize]
from time import time; from sys import stderr #you need these in order to print the time in stderr

#add the relative path for MiMeS/src
from sys import path as sysPath; sysPath.append('../src')	

from interfacePy.AxionMass import AxionMass #import the AxionMass class
from interfacePy.Axion import Axion #import the Axion class
from interfacePy.Cosmo import mP #import the Planck mass

def main():

    # AxionMass instance
    axionMass = AxionMass(r'../src/data/chi.dat',0,mP)

    # define ?$\maT^2$? for ?$T\leq T_{\rm min}$?
    TMin, chiMin=axionMass.getTMin(), axionMass.getChiMin()
    
    axionMass.set_ma2_MIN( lambda T,fa: chiMin/fa/fa )

    # define ?$\maT^2$? for ?$T\geq T_{\rm max}$?
    TMax, chiMax=axionMass.getTMax(), axionMass.getChiMax()

    axionMass.set_ma2_MAX( lambda T,fa: chiMax/fa/fa*pow(TMax/T,8.16))

    #in python it is more convenient to use relative paths
    inputFile="../UserSpace/InputExamples/MatterInput.dat" 

    ax = Axion(0.1, 1e16, 500, 1e-4, 1e3, 10, 1e-2, inputFile, axionMass, 
        1e-2, 1e-8, 1e-2, 1e-10, 1e-10, 0.85, 1.5, 0.85, int(1e7))

    ax.solveAxion()

    print("theta_i=",ax.theta_i,"\t\t\t\t","f_a=",ax.fa,"GeV\n","theta_osc~=",
            ax.theta_osc,"\t","T_osc~=",ax.T_osc,"GeV \n","Omega h^2=",ax.relic)

    #once we are done we should run the destructor
    del ax,axionMass

if __name__ == '__main__':
    _=time()
    main()
    print(round(time()-_,3),file=stderr)
\end{lstlisting}
	
	In order for this script to work, we should place it in a subdirectory os the root of \mimes. This is the only requirement, assuming that we have already compiled the shared libraries.

\section{Outlook}
We briefly introduced the basics of \mimes. We have discussed that it can be used to solve the axion/ALP EOM for various underline cosmologies and axion/ALP masses. We have also shown that \mimes is flexible and the user can change almost everything that it needs; \eg the plasma relativistic degrees of freedom, the convergence conditions of the integration, and underline the Runge-Kutta methods used.      

However, \mimes can be extended in the future, in order to make it even more general. For example, \mimes can be generalized to allow the user to consider different initial value of $\dot \theta$. \mimes can, conceivably, handle non-vanishing \rhs; \ie solve the "driven" dumped time-dependent pendulum. Another interesting addition would be to be able to compare against searches on the fly, by introducing lists of constraints from various experiments and observations.

%

\bibliography{refs}{}
\bibliographystyle{JHEP}                        

\end{document}